\documentclass[conference]{IEEEtran}
\IEEEoverridecommandlockouts

\usepackage{amsmath}
\usepackage{amssymb}
\usepackage{amsfonts}
\usepackage{algorithmic}
\usepackage{booktabs}
\usepackage{siunitx}
\usepackage{cite}
\usepackage{tikz}
\usetikzlibrary{arrows.meta,positioning,fit,backgrounds,calc}
\usepackage[hidelinks]{hyperref}
\usepackage{orcidlink}

\newcommand{\svc}[1]{\mbox{SVC-#1}}
\newcommand{\code}[1]{#1}
\begin{document}

\title{A Hybrid, Multi-Layered Pipeline for Phishing and\\
Threat Classification: Independently Validated URL and NLP\\
Engines with a Calibrated Multi-Channel Fusion Stage}

\author{%
\IEEEauthorblockN{Saifelden M.~Ismail\,\orcidlink{0009-0002-8867-6533},
Aser O.~Ibrahim,
and Omar A.~Mahmoud\,\orcidlink{0009-0003-9266-4102}}%
\IEEEauthorblockA{\textit{Department of Communications and Information Engineering}\\
\textit{Zewail City of Science and Technology}, Giza, Egypt\\
s-saifelden.ismail@zewailcity.edu.eg, s-aser.ibrahim@zewailcity.edu.eg,\\
s-omar.mahmoud@zewailcity.edu.eg}%
}

\maketitle

\begin{abstract}
Phishing is a multi-modal threat. We present a hybrid pipeline that scores each modality with
its own engine and fuses the results. Three engines are built, deployed, and independently
benchmarked: a four-stage URL stack (Domain Guard, lexical model, threat intelligence, and an
asymmetric L2 fusion sidecar); a generalization-hardened DistilBERT NLP classifier whose
held-out \emph{real}-phishing recall rises from \SI{0.8}{\percent} to \SI{87.3}{\percent}; and a
threat-intelligence synchronizer with end-to-end OpenTelemetry instrumentation confirming 1:1
message conservation. A decision-level fusion stage, characterized on a $10{,}677$-email
whole-system benchmark, reaches $F_1{=}0.914$ with a calibrated probabilistic-OR over URL,
header, and phishing-probability channels while reducing held-out real-spam false positives to
\SI{3.6}{\percent}. Because that benchmark uses proxy URL and header channels and an operating
point still needing recalibration, we present it as a preliminary integrated result. For
deployable detection, the limiting factor is how well a model generalizes, not how accurately
it scores data drawn from its own training distribution.
\end{abstract}


\section{Introduction}

Phishing remains the dominant initial-access vector for credential theft, business email
compromise, and malware delivery. A single message couples lure text, URLs, sender metadata, and
often an attachment, yet most detectors still reason over one modality at a time. A URL classifier
cannot read intent from prose; a text classifier cannot judge a freshly registered domain; a
reputation feed cannot recognize a lure it has never seen. Our premise is that classification
works better when independent, modality-specific signals are fused than when a single model is
asked to arbitrate the whole message.

Single-modality detectors fail in characteristic ways. A lexical URL model trained on a
mostly www-prefixed corpus reads a bare apex such as \textit{google.com} as phishing; exact-domain
threat intelligence misses subdomain rotation; and text classifiers overfit their training corpora,
as an early NLP checkpoint showed when it fell to \SI{0.8}{\percent} recall on held-out real
phishing. We address this with a layered design that keeps the modalities separate.

This paper documents the parts of the system that are realized and measured, and presents the
rest as supporting context. We make four contributions. First, a four-stage URL engine with an
embedded Domain Guard, a lexical first-level model, threat-intelligence lookup, and a
zone-based L2 fusion sidecar (Section~\ref{sec:l2}). Second, a generalization-hardened
DistilBERT email classifier whose held-out real-phishing recall rises from \SI{0.8}{\percent} to
\SI{87.3}{\percent} under a contrastive curriculum and shared canonicalization
(Section~\ref{sec:nlp}). Third, a production threat-intelligence synchronizer and end-to-end
instrumentation whose traces confirm 1:1 message conservation (Sections~\ref{sec:setup}
and~\ref{sec:ti}). Fourth, a decision-level fusion stage characterized on a
$10{,}677$-email whole-system benchmark, where a calibrated probabilistic-OR over URL, header,
and phishing-probability channels outperforms simpler combiners (Section~\ref{sec:decision}).

We are explicit throughout about what is benchmarked and how each measurement is obtained. The three modality engines are
each evaluated on held-out or representative data. The fusion benchmark relies on proxy URL and
header channels in place of production scores, and its threshold still needs recalibration on real
traffic, so we treat its numbers as a preliminary integrated result and not yet a production
accuracy claim. No real multi-channel phishing corpus with live URLs and full headers yet exists
against which fusion weights could be validated without circularity. This work was carried out as
a graduation project by the authors.

\section{Background and Related Work}

Our system is an event-driven collection of Go and Python microservices over a Kafka-API
broker, described in Section~\ref{sec:arch}. We position its design against three lines of
prior work: lexical URL detection, transformer-based email-text classification, and signal
fusion.

Lexical URL classification is a mature line of research, supplying established feature sets and public
corpora~\cite{chiew2019hefs,tamal2024dataset,mohammad2012assessment,potpelwar2025legitphish,prasad2024phiusiil}.
Our first-level URL model builds on that work and ranks candidates by Matthews correlation
coefficient~\cite{matthews1975} (Section~\ref{sec:urlml}). On top of that model we add a deterministic Domain Guard that
combines the Cisco Umbrella popularity ranking~\cite{ciscoumbrella}, Levenshtein edit
distance~\cite{levenshtein1966}, and the long-standing brand-in-subdomain abuse
pattern~\cite{edelman2003brand}. The second-level sidecar then fuses a calibrated
logistic-regression URL score~\cite{platt1999probabilistic} with a histogram
gradient-boosting model over operational host features~\cite{friedman2001greedy,ke2017lightgbm}
(Section~\ref{sec:l2}).

For email text we fine-tune DistilBERT~\cite{sanh2019distilbert}, following other
lightweight-transformer phishing detectors~\cite{altan2025dualpath,safran2025phishinggnn,atanda2026lightweight,tawfik2025xfphishbert},
and we harden the model with fast gradient-method adversarial training and character-level
noise augmentation~\cite{goodfellow2015adversarial,sajad2026robust}. The multi-task head
arrangement follows the multi-task learning
literature~\cite{ibrahim2024mtl,kendall2018uncertainty} and uses focal loss~\cite{lin2017focal}
for the imbalanced classification task; the intent taxonomy and the multi-source corpus
assembly draw on prior codebook and corpus work~\cite{mendes2025meajor,saka2024codebook}, and
the model is exported through ONNX with quantization studied
explicitly~\cite{neshaei2024quantization}. We use three classes, legitimate,
spam, and phishing, where most detectors use only two, because the spam-versus-legitimate confusion is
the dominant failure mode of binary and LLM classifiers~\cite{toth2025phish}
(Section~\ref{sec:nlp}).

Across these components, URL reputation, sender reputation, and textual intent are scored by
separate layers and recombined at the fusion stage of Section~\ref{sec:decision}.

\section{System Architecture}
\label{sec:arch}

We present the system as designed, but we distinguish clearly between the components that are
realized and measured and those that are only specified. Three services are deployed as fully
implemented, independently benchmarked engines: the URL analysis service \svc{03}, the NLP analysis service
\svc{06}, and the threat-intelligence synchronizer \svc{11}. The aggregator \svc{07} and the
decision engine \svc{08} are implemented and exercised on the whole-system benchmark of
Section~\ref{sec:decision}. The remaining services are specified and run as transport stubs
that exercise the end-to-end backbone.

\begin{figure}[t]
\centering
\begin{tikzpicture}[
  font=\scriptsize,
  node distance=1.8mm,
  svc/.style={draw, rounded corners=1pt, align=center, inner sep=1.5pt, minimum width=14mm, minimum height=4.2mm},
  real/.style={svc, very thick},
  stub/.style={svc, dashed},
  topic/.style={midway, font=\tiny\ttfamily},
  >=Latex
]
\node[svc] (s1) {\svc{01}\\Ingest};
\node[svc, below=of s1] (s2) {\svc{02}\\Parser};
\node[real, below left=3.5mm and -2mm of s2] (s3) {\svc{03}\\URL};
\node[stub, right=1.6mm of s3] (s4) {\svc{04}\\Header};
\node[stub, right=1.6mm of s4] (s5) {\svc{05}\\Attach};
\node[real, right=1.6mm of s5] (s6) {\svc{06}\\NLP};
\node[stub, below=6mm of s4] (s7) {\svc{07}\\Aggregator};
\node[stub, below=of s7] (s8) {\svc{08}\\Decision};
\node[stub, below left=3mm and 0mm of s8] (s9) {\svc{09}\\Notify};
\node[stub, below right=3mm and 0mm of s8] (s10) {\svc{10}\\API};
\node[real, right=8mm of s6] (s11) {\svc{11}\\TI Sync};

\draw[->] (s1) -- node[topic, right] {emails.raw} (s2);
\draw[->] (s2) -- (s3);
\draw[->] (s2) -- (s4);
\draw[->] (s2) -- (s5);
\draw[->] (s2) -- (s6);
\draw[->] (s3) -- (s7);
\draw[->] (s4) -- (s7);
\draw[->] (s5) -- (s7);
\draw[->] (s6) -- (s7);
\draw[->] (s7) -- node[topic, right] {emails.scored} (s8);
\draw[->] (s8) -- (s9);
\draw[->] (s8) -- (s10);
\draw[->, dotted] (s11) to[bend right=20] (s3);
\end{tikzpicture}
\caption{Processing topology of the system. Thick solid boxes (\svc{03}, \svc{06}, \svc{11})
are the independently benchmarked engines; dashed boxes are designed or auxiliary services.
\svc{11} feeds the Valkey threat-intelligence cache consumed by \svc{03}. All Kafka topics are
keyed by the email identifier.}
\label{fig:topology}
\end{figure}
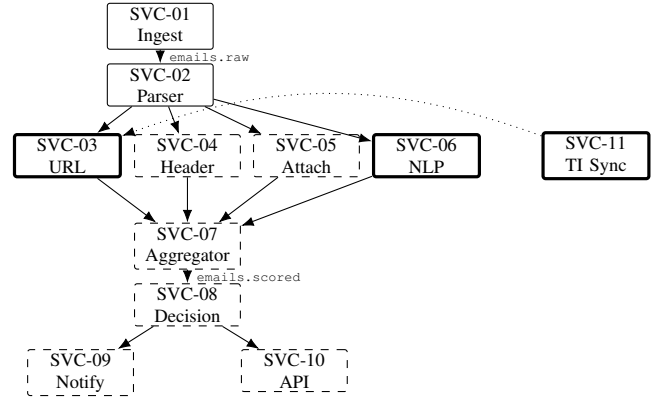

Inbound mail enters at the ingest service \svc{01}, which exposes five source adapters and
deduplicates on the pair of organization and message identifiers. The parser \svc{02} then
reads the MIME structure, extracts URLs, hashes attachments with three digests, computes
Shannon entropy and a magic-byte file type, and emits an analysis plan listing the scores it
expects downstream. Four analyzers next score independent modalities in parallel: the URL
engine \svc{03}, the header analyzer \svc{04}, the attachment analyzer \svc{05}, and the text
engine \svc{06}. Each publishes its own score envelope. The aggregator \svc{07} gathers these
behind a synchronization barrier, and the decision engine \svc{08} combines them, applies a rule
engine, maps the combined score to a verdict, fingerprints the campaign, and persists the result in a
single transaction. The synchronizer \svc{11} refreshes the threat-intelligence store out of
band (Fig.~\ref{fig:topology}).

The twelve Kafka topics are all keyed by the email identifier, so every message about one email
co-partitions within a topic and stays in order. Payloads are JSON carrying a versioned
envelope, and delivery is at-least-once with idempotent producers.

The score envelopes emitted by the realized engines are well-defined. The URL score message
carries, beside its integer score, whether threat intelligence matched, whether the guard
fired, the fused deployment probability, and the machine-learning verdict; the per-email URL
score is the maximum over the individual links. The NLP score message adds intent, urgency,
impersonation, and deception facets.

Persistence spans roughly twenty-three PostgreSQL tables, with the email table range-partitioned
by month. Threats enriched from observed email traffic are kept in a separate table from the
low-cost bulk indicators that the synchronizer writes (Section~\ref{sec:ti}), so expensive
enrichment never mixes with feed data. Verdicts are append-only. Per-organization row-level
security is specified but not yet enforced.

The full system, including operational binaries and documentation, is released publicly at
\url{https://github.com/XHCFS/cybersiren}.

\section{Datasets}
\label{sec:datasets}

Because the engines are trained and evaluated separately, the data that feeds each one merits
separate description. This section documents the two largest supervised corpora, the
URL corpus behind the first-level URL model and the multi-source email corpus behind the NLP
model, together with how each was assembled, de-duplicated, split, and labeled, and what biases
each carries. The operational enrichment data for the fusion sidecar and the whole-system
benchmark corpus are described alongside the components that use them, in
Sections~\ref{sec:l2} and~\ref{sec:decision}.

\subsection{The URL corpus}
The URL model learns from raw URL strings, so we only considered datasets that expose the
original URL text; any dataset exposing only pre-extracted or discretized features was excluded. Two public
datasets met it: PhiUSIIL~\cite{prasad2024phiusiil}, with $235{,}795$ URLs, and
LegitPhish~\cite{potpelwar2025legitphish}, with $101{,}219$ manually verified URLs. Four other
candidates were rejected for specific reasons, which we document: two
encode their features as ternary values with no raw URL column, and two more carry large blocks
of exact duplicates, including one dataset that is \SI{47.6}{\percent} duplicate rows. Both
chosen datasets natively label legitimate as one and phishing as zero, which we invert to the
conventional scheme of zero for legitimate and one for the positive phishing class.

Merging the two sets required several cleaning steps. We dropped rows with a null URL or label and rows whose URL
was shorter than ten characters or contained no dots, then de-duplicated case-insensitively,
keeping the PhiUSIIL copy on any conflict because it is the larger and more recent source. This
removed $37{,}706$ rows. The resulting corpus holds $299{,}306$ URLs, of which
\SI{45.2}{\percent} are legitimate and \SI{54.8}{\percent} phishing, with no null or infinite
feature values. We split it $70/15/15$ with stratification, and the held-out test partition of
$44{,}896$ URLs was scored exactly once. Three lookup tables, all built from the Cisco Umbrella
top-million list~\cite{ciscoumbrella}, support the feature extractor: a per-character probability table over a corpus
of roughly twenty-one million characters, a table of legitimacy probabilities for $1{,}319$
top-level domains, and a sixteen-word list of security-sensitive terms.

Two biases in this corpus should be noted explicitly, because both shape how the model is
deployed. The data spans roughly 2015 to 2023, so concept drift is expected as phishing tactics
evolve. More importantly, the near-even class balance is markedly different from the roughly
one-percent phishing prevalence of real inbound traffic, which is precisely why the deployed
model runs at a conservative threshold and is scheduled for recalibration on production data. The
corpus is also dominated by Latin-script URLs, leaving internationalized domains
underrepresented.

\subsection{The email corpus}
The NLP model is trained on six real corpora that together cover the three classes it must
distinguish. The foundation is the Biggest Spam--Ham--Phish Kaggle corpus of
$365{,}448$ messages~\cite{akshatsharma2023spamhamphish}, the only source with native
three-class labels; it carries $12{,}477$ internal label conflicts, which we resolve by
majority vote. A modern educational-phishing set~\cite{tanvirahmed2024eduphish} and a corpus of
human- and machine-generated phishing emails~\cite{greco2024llmphish} supply contemporary and
AI-written lures; that study releases and evaluates the Kaggle corpus we ingest. The Nazario~\cite{nazario2005phishing} and Nigerian-419~\cite{radev2008nigerian}
collections contribute real-world phishing from 2004 onward. Finally, a large aggregation of
seven historical spam corpora~\cite{puyang2025seven}, including the TREC, CEAS, Enron,
SpamAssassin, and LingSpam sets,
supplies the dedicated spam class; during profiling we corrected its labeling, recognizing that
its positive label denotes spam, not phishing, since every sub-corpus in it is historically a
spam collection. Four further candidates were excluded after inspection, among them two that
proved synthetic or near-duplicate at the level of a few dozen unique texts, and one that had
been pre-stemmed beyond use.

All sources are mapped to the integer scheme of zero for legitimate, one for spam, and two for
phishing, which is distinct from the platform's string verdict enumeration. We then
de-duplicate in two passes, first by an exact hash of the normalized text and then by
MinHash locality-sensitive hashing~\cite{broder1997minhash} at a Jaccard threshold of $0.80$,
keeping the
longest sample in each near-duplicate cluster. Splitting happens at the level of whole
campaigns, never at the level of individual messages: every message is fingerprinted by its
subject and first hundred characters, and a grouped
shuffle split sends all variants of a campaign to the same partition, with assertions that no
fingerprint and no more than a five-percent class-ratio deviation crosses partitions. Random
splitting would let a campaign template appear in both training and test and inflate every
metric, so this step is essential. The target composition is $40/40/20$ across legitimate,
phishing, and spam: phishing is over-represented so the model learns its decision boundary
thoroughly, and spam is held at a fifth of the corpus specifically to prevent the spam-as-phishing
confusion that dominates three-class evaluations~\cite{toth2025phish}.

The change that separated the model's first and deployed versions was not the data sources but
how they were assembled, and we describe it here because it is as much a dataset decision as a
training one. The real phishing collections were moved roughly \SI{85}{\percent} into training,
with a \SI{15}{\percent} held-out real-phishing test split of $369$ messages reserved as the
generalization metric we rely on. Every source was capped so that no single corpus's vocabulary
could dominate the gradient. Hard phishing patterns were paired with near-identical legitimate
twins, which forces the model to rely on the intent that distinguishes them and not on a surface
phrase.
Character-level noise was added to about a quarter of the phishing and a twelfth of the
legitimate messages, and leetspeak substitution to about a sixth of the phishing, yielding a
training set of roughly $188{,}000$ rows. As with the URL corpus, the main biases
warrant explicit statement: the $40/40/20$ balance is unlike the roughly $85/5/10$ mix of real
inboxes and so requires threshold recalibration, the tokenizer is English-only, and the
$256$-token input window truncates unusually long messages.

\section{Experimental Setup and Observability}
\label{sec:setup}

Before turning to the engines, we fix the evaluation protocol and the vocabulary used to report
it, since the credibility of the reported figures depends on being precise about how each was
obtained. The guard and fusion-sidecar latency figures were measured on an Intel
i7-10750H. The NLP model was benchmarked as deployed, running on CPU under ONNX Runtime with
full graph optimization; an optimization cache reduces its cold start from one to two minutes
down to a few seconds. The URL fusion sidecar runs as an out-of-process Python service that the
Go engine calls over HTTP, which keeps the Go binaries small and lets the models be reloaded
without restarting the analysis service.

Three phrases recur in the results and each has a concrete meaning here. A \emph{held-out}
measurement is taken on a partition that the model never saw during training and, where the
split is campaign-aware, never saw even a sibling of. A \emph{production-like} figure is one
measured on the most demanding realistic subset available, not the least demanding, such as subtle generic-domain
phishing that carries no brand cue, and we report it as a lower bound, not a best-case figure. A
\emph{snapshot} is a single point-in-time capture, such as one day's fresh phishing feed; a
snapshot number has a wide confidence interval and is always paired with the lower bound it
should be read against. The primary metrics follow the same conventions throughout: Matthews
correlation coefficient~\cite{matthews1975} as the primary metric for the URL model because it accounts
for all four
confusion-matrix quadrants under class imbalance, area under the ROC curve and false-positive
and false-negative rates for the probabilistic models, macro-averaged $F_1$ and macro Matthews
correlation coefficient~\cite{matthews1975} for the three-class NLP task, and expected
calibration error for probability quality.

Every service is instrumented along three pillars: Prometheus metrics, OpenTelemetry traces
exported to Jaeger, and structured logs, with the host interfaces for Prometheus, Grafana, and
Jaeger each on their own port. Trace context propagates across services through Kafka headers, so a single trace spans the
whole pipeline, and three shared counters for messages consumed, messages produced, and
per-stage processing time let us verify message conservation end to end. The full setup is
reproducible from the demonstration stack.

\section{Low-Latency URL Model (L1)}
\label{sec:urlml}

The first-level URL model maps a raw URL to a phishing probability from structural features
alone, with no network calls at inference. It composes a deterministic extractor
$\phi:\Sigma^{*}\rightarrow\mathbb{R}^{30}$ with a gradient-boosted ensemble
$g:\mathbb{R}^{30}\rightarrow[0,1]$, so the model is $f(u)=g(\phi(u))$, and the deployed
model is an XGBoost~\cite{chen2016xgboost} classifier at a threshold of $0.85$. Its training data is the merged
$299{,}306$-URL corpus of Section~\ref{sec:datasets}, balanced $45.2$ to $54.8$ between
legitimate and phishing.

The feature vector has thirty dimensions. Twenty-eight are structural and lexical signals,
including URL and hostname lengths, dot and hyphen counts, subdomain statistics, the Shannon
entropy of the URL and of the domain, the character-probability and continuation-rate scores
derived from the PhiUSIIL work~\cite{prasad2024phiusiil}, a top-level-domain legitimacy
probability, and several binary flags. The remaining two are legitimacy anchors, a minimum
brand edit distance and a flag for membership in the top-million domain list, which give the
model a direct popularity and brand signal for deep-path or high-entropy URLs. Two classically
promoted features, the presence of an IP-address host and a double slash in the path,
contributed no splits at all and were pruned. Importance is dominated by the entropy and
probability-table features; the lexical flags that the literature usually emphasizes matter less
(Table~\ref{tab:urlimp}): the top five features carry \SI{63.1}{\percent} of cumulative
importance, and the character-probability feature alone carries \SI{15.5}{\percent}.

\begin{table}[t]
\caption{Top-10 L1 features by deployed-model split count.}
\label{tab:urlimp}
\centering
\footnotesize
\begin{tabular}{@{}rlrr@{}}
\toprule
Rank & Feature & Splits & Cumul.\ \% \\
\midrule
1 & \code{url\_char\_prob}          & 2{,}697 & 15.5 \\
2 & \code{entropy\_domain}          & 2{,}293 & 28.7 \\
3 & \code{entropy\_url}             & 2{,}244 & 41.6 \\
4 & \code{tld\_legit\_prob}         & 2{,}081 & 53.6 \\
5 & \code{char\_continuation\_rate} & 1{,}659 & 63.1 \\
6 & \code{url\_length}              & 1{,}171 & 69.9 \\
7 & \code{hostname\_length}         & 844     & 74.7 \\
8 & \code{path\_length}             & 708     & 78.8 \\
9 & \code{avg\_subdomain\_length}   & 692     & 82.8 \\
10& \code{pct\_numeric\_chars}      & 413     & 85.2 \\
\bottomrule
\end{tabular}
\end{table}

Model selection used a controlled comparison. Ten classifiers were trained under the same $70/15/15$
stratified split and ranked by Matthews correlation coefficient~\cite{matthews1975}, which we
prefer to accuracy or
$F_1$ here because it weighs all four confusion-matrix quadrants under the corpus's class
imbalance. One minor but useful fix emerged during this process: a fallback top-level-domain
parser failed on the bare domains in the Cisco Umbrella list~\cite{ciscoumbrella}, and installing
a dedicated parser
not only fixed the extractor but raised the top-level-domain legitimacy feature to fourth in
importance.

The primary results are as follows. On the $44{,}896$-URL held-out test set, the best model, LightGBM~\cite{ke2017lightgbm} at the
default $0.50$ threshold, reaches a Matthews correlation coefficient of $0.99645$. Its accompanying metrics are
an accuracy of \SI{99.824}{\percent}, precision of \SI{99.919}{\percent}, recall of
\SI{99.760}{\percent}, and an $F_1$ of \SI{99.839}{\percent}. Its area under the ROC curve is
$0.99929$ and its log loss $0.01026$, at a false-positive rate of \SI{0.099}{\percent} and a
false-negative rate of \SI{0.240}{\percent}, corresponding to a confusion matrix of $20{,}275$
true negatives, $20$ false positives, $59$ false negatives, and $24{,}542$ true positives. The
deployed artifact is XGBoost~\cite{chen2016xgboost}, not LightGBM, and it runs at the
conservative $0.85$ threshold;
its embedded metrics, an accuracy of $0.9658$, an $F_1$ of $0.9683$, and a Matthews coefficient
of $0.9333$, are lower precisely because the higher threshold trades recall for precision; we
state the two operating points separately to keep them from being conflated.
Feature extraction runs at roughly sixty thousand URLs per second and single-URL inference in
about $1582$ microseconds.

At runtime the first-level model executes only after the Domain Guard (Section~\ref{sec:l2}). In
the user-facing scan endpoint, the handler deliberately forces the first-level score to zero, so
the verdict is set by the threat-intelligence match and the fusion sidecar; the score-threshold
branches run only in the batch pipeline. The first-level score is still computed and returned
for observability. Residual limitations are the corpus biases noted in Section~\ref{sec:datasets},
which require threshold recalibration after deployment.

\section{Domain Guard and L2 Fusion}
\label{sec:l2}

The first-level model of Section~\ref{sec:urlml} sits inside a four-stage URL pipeline:
Domain Guard, first-level model, threat-intelligence lookup, and L2 fusion sidecar, with each
earlier stage able to short-circuit the later ones. This section covers the guard and sidecar;
the decision engine that consumes their output is described in Section~\ref{sec:decision}.

\subsection{Domain Guard}
The guard runs first and is written in Go. It begins by deriving the registrable apex of a URL
through the Public Suffix List~\cite{publicsuffix}, and then applies three signals in order. A ten-thousand-entry
allowlist of the most popular Cisco Umbrella domains~\cite{ciscoumbrella}, embedded into the
binary at build time as a hash set for constant-time lookup, returns a legitimate verdict on a
hit and skips the rest of the pipeline. Failing that, a Levenshtein-distance-one typosquat
check~\cite{levenshtein1966} compares the apex against the same ten thousand references; a
length-gap test discards any candidate differing by more than one character before the
edit-distance computation runs, and a match yields a phishing verdict. Finally, a
brand-in-subdomain scan~\cite{edelman2003brand} checks non-apex labels of at least four
characters against twenty hand-curated brand substrings and likewise returns phishing on a hit.
The guard is the reason the downstream fusion weights can be aggressive without false positives
on well-known domains, because those domains never reach the sidecar at all. Measured on an
i7-10750H, an allowlist hit resolves in about $43$ nanoseconds, a worst-case full-scan
typosquat check in about $51$ microseconds, a brand-in-subdomain hit in about $157$ nanoseconds,
and apex extraction in under a microsecond.

\subsection{The two fused models}
When the guard does not fire and threat intelligence has not already returned a high-confidence
verdict, the sidecar computes two probabilities and fuses them.

The first is a character-level URL probability. A character $n$-gram model over $n\in[2,4]$
feeds a hashing vectorizer with $2^{18}$ buckets, which means there is no fitted vocabulary and
the model stays stateless under data drift. Its output drives an L2-regularized logistic
regression with $C{=}2.0$, balanced class weights, the \emph{liblinear} solver, and a
three-thousand-iteration cap. Because logistic regression emits a native sigmoid probability, this score is already
calibrated for linear combination and needs no separate Platt-scaling
calibrator~\cite{platt1999probabilistic} of the kind used for margin-based classifiers. Inputs are normalized to ASCII lowercase with
percent-decoded paths and punycode-decoded hosts. An early version of this model was seriously
flawed: it scored an area under the curve of only
$0.757$, the product of an overfit augmentation scheme and direct data leakage, since its phishing
augmentation was drawn from the very benchmark used to test it. The deployed third version was
retrained on $304{,}181$ rows, comprising the merged base corpus plus a realistic fresh
OpenPhish augmentation~\cite{openphish} and a hard-benign augmentation, and removing the leakage
lifted held-out
performance to the figures reported below.

The second is an operational probability from a histogram gradient-boosting
classifier~\cite{friedman2001greedy,ke2017lightgbm} over forty-four features of the live host:
twenty-eight raw enrichment fields spanning autonomous-system, geolocation, TLS, HTTP, and
WHOIS data, and sixteen derived features. Three of the derived features describe page-level
structure that the current training data predates, so they are trained as missing values; the
classifier handles them natively by learning no split on a fully absent column. High-cardinality
categorical fields are hashed into four thousand numeric buckets.
The base operational training corpus comprises $100{,}000$ labeled URLs with balanced phishing
and benign classes, materialized as static live-enrichment feature rows under stratified
train, validation, and test splits in \texttt{ready\_operational}; the held-out test split
contains $15{,}000$ URLs ($7{,}500$ per class).
Two synthetic injections into
the training data serve a substantive purpose and are not cosmetic. Fifteen hundred platform rows teach the
model that hosting on an ephemeral platform such as Vercel, Netlify, or GitHub Pages is not by
itself a reliable indicator of phishing, by pairing benign brand-authentication and developer-deployment pages against
phishing platform pages so the model learns to discriminate on brand-in-subdomain count, title
mismatch, and domain age, with the platform itself carrying little weight. Four hundred benign shortener rows
do the same for bit.ly, t.co, and similar domains, which are absent from phishing feeds and so
would otherwise have no prior at all.

\subsection{Fusion decision function}
The fusion combines the URL probability $\textit{url}_p$ and the operational probability
$\textit{op}_p$ into a deployment probability $\textit{deploy}_p$ through an asymmetric,
zone-based rule. Three boolean masks select the zones:
\begin{align}
\textsf{cdn} &\equiv (\textit{op}_p < 0.01),\\
\textsf{high\_op} &\equiv (\textit{op}_p > 0.60)\wedge(\textit{url}_p \ge 0.05),\\
\textsf{high\_op\_guard} &\equiv (\textit{op}_p > 0.60)\wedge(\textit{url}_p < 0.05).
\end{align}
The design follows from the distinct error modes of the two models. The URL probability
over-flags keyword-heavy small-business banking domains, while the operational probability
over-flags CDN-fronted phishing whose infrastructure looks clean. Each model is therefore
allowed to dominate only where it is reliable. \eqref{eq:fusion} gives the rule,
evaluated top to bottom so that the first matching clause wins, where $\textsf{short}$ marks a
known URL shortener; a deployment threshold of $0.50$ then yields the binary verdict.
\begin{equation}
\label{eq:fusion}
\textit{deploy}_p =
\begin{cases}
0.10\,\textit{url}_p + 0.90\,\textit{op}_p, & \textsf{short} \wedge \textit{url}_p \le 0.95 \\[2pt]
0.65\,\textit{url}_p + 0.35\,\textit{op}_p, & \textsf{cdn} \\[2pt]
0.60\,\textit{url}_p + 0.40\,\textit{op}_p, & \textsf{high\_op\_guard} \\[2pt]
0.25\,\textit{url}_p + 0.75\,\textit{op}_p, & \textsf{high\_op} \\[2pt]
0.60\,\textit{url}_p + 0.40\,\textit{op}_p, & \text{otherwise (normal)}
\end{cases}
\end{equation}
The shortener override down-weights the URL probability because the shortener apex is itself
uninformative; the exception, retained as the normal blend, is when the URL model is more than
ninety-five percent confident the shortener domain is itself malicious. The high-operational
guard exists because legitimate platform deployments land at a URL probability of roughly
$0.001$ to $0.026$, whereas real phishing in the same operational zone carries a URL
probability of at least $0.06$; falling back to the normal blend there suppresses false
positives without suppressing true positives. One revision subtracted a zone instead of adding
one, the only such change, and it too was driven by evidence: an earlier ``established-domain'' zone over-weighted the
operational probability for very clean-looking infrastructure, and a benchmark attributed
roughly eighty extra false negatives per run to it, so it was deleted.

\subsection{Sidecar and integration}
The Python sidecar exposes a small HTTP surface for health, batch scoring, single-URL scoring,
and a feature-bypass path that the Go enricher uses when it has already computed the operational
features. A Go detector with a default threshold of $0.50$ calls it through a five-minute,
ten-thousand-entry cache keyed on the full URL. Keying on the apex alone had allowed distinct paths
on the same domain to collide, and keying on the full URL eliminated these collisions. The sidecar is consulted whenever
threat intelligence did not match or matched below a risk score of $80$, which turns a
low-confidence intelligence hit into a hint the fusion can confirm or reject, stopping short of
a standalone verdict. A guard at the network boundary rejects connections to non-public addresses
to limit server-side request forgery, and the sidecar is fail-open: if it is unreachable the
guard and intelligence verdict stands and the machine-learning fields are simply omitted.

\subsection{Measured performance}
Table~\ref{tab:l2dr} reports detection rate across six phishing corpora and false-positive rate
on a clean held-out benign split, with the sidecar measured in isolation and the production
allowlist deliberately not applied. Detection ranges from every scored sample on a fresh, unseen
OpenPhish snapshot down to \SI{74.7}{\percent} on subtle generic-domain phishing that carries no
brand cue in the URL, which is the realistic lower bound for production-like traffic. The clean
held-out benign false-positive rate is \SI{1.3}{\percent}. The much higher rates seen on the
hard-benign corpora reflect well-known domains that the production allowlist short-circuits
before the sidecar is ever consulted, so the clean held-out figure is the representative one.
Per model, the third-version URL probability attains a held-out area under the curve of $0.974$
with a benign false-positive rate of \SI{0.6}{\percent} at threshold $0.50$, and the operational
model attains validation and test areas under the curve of $0.99985$ and $0.99968$.

\begin{table}[t]
\caption{L2 sidecar detection rate (DR) by corpus; clean benign FPR.}
\label{tab:l2dr}
\centering
\footnotesize
\begin{tabular}{@{}lrr@{}}
\toprule
Corpus & Scored & DR \\
\midrule
Fresh OpenPhish (unseen at training)      & 286 & \textbf{100.0\%} \\
Mixed phishing (OpenPhish/PhishTank~\cite{phishtank}/manual)& 307 & 97.5\% \\
Live ground-truth phishing                 & 75  & 98.5\% \\
Subdomain-chain adversarial                & 321 & 89.4\% \\
LegitPhish/PhiUSIIL subtle phishing        & 150 & 74.7\% \\
Error-analysis false negatives             & 61  & 16.4\% \\
\midrule
LegitPhish/PhiUSIIL benign (clean held-out)& 150 & 1.3\% (FPR) \\
\bottomrule
\end{tabular}
\end{table}

\section{Threat-Intelligence Synchronization (\svc{11})}
\label{sec:ti}

The synchronizer is a standalone scheduled Go service that sits off the Kafka path. It ingests
external feeds into PostgreSQL, refreshes the materialized views built over them, and rebuilds a
Valkey domain-blocklist cache that the URL engine queries in sub-millisecond time. A
representative live run held $2{,}678$ active indicators across four feeds and roughly four
hundred cached domain keys, and a full cycle over those feeds completes in about eight seconds.
Table~\ref{tab:feeds} lists the five feeds: PhishTank~\cite{phishtank},
OpenPhish~\cite{openphish}, URLhaus~\cite{urlhaus}, ThreatFox~\cite{threatfox}, and
MalwareBazaar~\cite{malwarebazaar}, with their fixed risk and confidence priors and the counts
observed in that run.

Two design choices shape how the store behaves. Indicators upsert on a per-feed key of feed,
type, and value, so the same indicator reported by two feeds yields two rows and the feed count
itself becomes a corroboration signal; on conflict the risk score is merged by taking the
greater of the two, and indicators no longer seen in a sync are marked inactive instead of being
deleted, which lets downstream services tell ``currently blocklisted'' from ``was
blocklisted.'' A deliberate schema separation routes bulk feed indicators to the indicator
table and malware hashes to the attachment library, which keeps the costly per-host enrichment
that populates the email-observed threat table off the low-cost bulk feed data entirely.

\begin{table}[t]
\caption{Ingested TI feeds (live-run counts shown where available).}
\label{tab:feeds}
\centering
\footnotesize
\begin{tabular}{@{}llrrr@{}}
\toprule
Feed & Types & Risk & Conf. & Live \\
\midrule
PhishTank      & URL                & 90 & 0.95 & n/a \\
OpenPhish      & URL                & 85 & 0.80 & 300 \\
URLhaus        & URL                & 80 & 0.75 & 1{,}946 \\
ThreatFox      & URL/dom/IP/hash    & \multicolumn{2}{c}{conf.-level} & 432 \\
MalwareBazaar  & hash $\rightarrow$ lib. & 90 & 0.95 & 30 \\
\bottomrule
\end{tabular}
\end{table}

\section{NLP Email Classification (\svc{06})}
\label{sec:nlp}

The NLP engine classifies email text into legitimate, spam, and phishing, and emits a content
risk score, intent labels, an urgency score, and an obfuscation flag. It is a fine-tuned
DistilBERT model served through ONNX Runtime behind a FastAPI process. The shipped checkpoint,
cycle-12, is generalization-hardened; an earlier checkpoint posted strong same-distribution
metrics yet collapsed on held-out real phishing (Table~\ref{tab:nlpres}).

\subsection{Model and preprocessing}
The backbone is the base uncased DistilBERT, with roughly sixty-six million
parameters~\cite{sanh2019distilbert}, carrying three heads on a shared encoder. The
classification head is a softmax over the three classes trained with focal loss at
$\gamma=2.0$~\cite{lin2017focal}; an eleven-class multi-label intent head, whose taxonomy
draws on qualitative phishing codebook work~\cite{saka2024codebook}, uses per-class sigmoids;
and a scalar head predicts urgency.
During fine-tuning, intent and urgency losses are masked on every sample in the shipped
recipe: the only corpora that carried those annotations were excluded from training because
profiling showed they were synthetic template data and not representative campaigns, so the
auxiliary heads ride on the shared encoder without direct supervision in cycle-12.
At serving time, intent labels and the urgency score are produced by the deployment keyword
rules (regular expressions over the eleven-class taxonomy and a scaled count of urgency
phrases) instead of by the auxiliary heads, which remain inactive until annotated training
data is available.
The three losses are balanced automatically by homoscedastic uncertainty
weighting~\cite{kendall2018uncertainty}, which keeps the easy classification task from
dominating the auxiliary heads. The input is the subject
and body concatenated and tokenized to $256$ tokens with a head--tail split of $64$ leading and
$190$ trailing tokens, so that both the opening hook and the footer call-to-action survive
truncation.

Preprocessing is the design choice that matters most, and it lives in a single module shared by
training and serving so the two paths can never drift apart. Its eleven steps first strip URLs
and bare email addresses, because their reputation is the job of the URL and header engines and
removing them pushes the model toward learning intent and away from memorizing link strings. The
remaining steps canonicalize adversarial surface forms: Unicode compatibility normalization,
cross-script homoglyph folding, leetspeak folding, the rejoining of letter-spaced words, and the
removal of zero-width characters. Because canonicalization maps a known attack back onto the
clean distribution at inference time, most character- and encoding-level evasions are defended
without retraining. A new encoding calls for a new normalizer, not a new training cycle.

\subsection{Training and generalization}
Fine-tuning uses AdamW~\cite{loshchilov2019adamw} with a learning rate of $2\times10^{-5}$, a warmup ratio of $0.1$, a
batch size of $32$, three epochs, weight decay $0.01$, and a head dropout of $0.3$. Robustness
comes from fast gradient-method adversarial training~\cite{goodfellow2015adversarial,sajad2026robust},
which perturbs the input embeddings along the normalized loss gradient,
$\delta=\epsilon\,\nabla_e L_\text{clean}/\lVert\nabla_e L_\text{clean}\rVert_2$ with
$\epsilon=0.01$, and trains on the sum of the clean and adversarial losses,
$L_\text{total}=L_\text{clean}+\lambda L_\text{adv}$ at $\lambda=0.5$. It is reinforced by
character-noise augmentation on about a quarter of the phishing and a twelfth of the legitimate
messages and leetspeak substitution on about a sixth of the phishing.

What turned the overfit baseline into a deployable model was the dataset-side redesign of
Section~\ref{sec:datasets}: a held-out real-phishing test split, per-source caps, and paired
contrastive buckets. We verify robustness with a metamorphic invariant in place of a frozen
attack list: any semantics-preserving transform of a phishing message the model already catches
must not let it evade, and a registry of twelve such transforms is applied automatically to a
corpus of caught phishing. Adding a future attack class is one transform function, applied to
every base case.

\subsection{Results}
Table~\ref{tab:nlpres} contrasts the overfit first version with cycle-12. Same-distribution
metrics are strong for both, but only cycle-12 generalizes on held-out real phishing, passes a
sixty-four-axis deployability gate, and records zero evasions across the metamorphic battery and
a never-seen probe set.

The residual errors concentrate on understated business lures whose disambiguating signal is the URL
and sender that the model deliberately removes. Pushing recall on these cases costs an equal
amount of legitimate precision (a text-only Pareto frontier), so the architecture-aware decision
is to defer them to the URL, sender, and fusion layers.

\begin{table}[t]
\caption{NLP results: overfit v1 vs.\ shipped cycle-12 (held-out test except where noted).}
\label{tab:nlpres}
\centering
\footnotesize
\begin{tabular}{@{}lrr@{}}
\toprule
Metric & v1 (overfit) & v2 cycle-12 \\
\midrule
Macro $F_1$               & 0.987 & \textbf{0.979} \\
Macro MCC                 & 0.980 & \textbf{0.969} \\
Phishing recall           & 0.986 & \textbf{0.978} \\
Legitimate FPR            & 0.008 & \textbf{0.0118} \\
Held-out \emph{real}-phishing recall & \textbf{0.008} & \textbf{0.873} \\
64-axis deployability gate & fail & \textbf{PASS} \\
Metamorphic battery (12)   & evades & \textbf{0 evasions} \\
Novel probe (8 phish/6 legit) & n/a & \textbf{0 FN / 0 FP} \\
\bottomrule
\end{tabular}
\end{table}

\subsection{Serving and quantization}
The service runs as two processes in one container, a Go wrapper on one port and a Python
inference engine on another (Section~\ref{sec:setup}). The content risk score is
threshold-independent, the rounded phishing probability scaled to a hundred, with a tuned
promotion floor of $0.24$ on the phishing probability for borderline cases. We did not assume
quantization would be beneficial; we evaluated it, following prior work on transformer
quantization robustness~\cite{neshaei2024quantization}.
An early INT8 attempt failed because of an export-path bug, and a corrected re-run on cycle-12
measured the actual trade-off: dynamic INT8 is four times smaller and $1.83$ times faster and
identical on confident cases, but it perturbs borderline scores by up to seventy-seven points,
and static INT8 eliminates confident phishing detections entirely. Because the score feeds an aggregated
verdict and must stay stable on borderline emails, the shipped precision is faithful 32-bit
floating point of about $266$ megabytes, exact to the PyTorch model with a maximum logit
difference of zero. Measured CPU latency is $9.8$ milliseconds at the median and $11.3$
milliseconds at the 95th percentile, an order of magnitude inside the service budget.

\section{Decision-Level Fusion and the Whole-System Benchmark}
\label{sec:decision}

The three engines of the preceding sections each emit a per-modality score. The aggregator
\svc{07} and decision engine \svc{08} (Section~\ref{sec:arch}) turn those scores into one
verdict: a synchronization barrier collects channel scores, the blender fuses them, a rule engine
and verdict bands map the result, and a campaign fingerprint groups template variants by sender,
URL, subject hash, intent, and 64-bit SimHash~\cite{manku2007simhash}.

\subsection{The fusion function}
The blender is configurable, and the choice of mode is the main design decision. The
production default is a calibrated probabilistic-OR: each channel's score is mapped to a
calibrated phishing probability, and the channel probabilities are combined as a noisy-OR~\cite{pearl1988}
so
that one confident channel is enough to raise the verdict. Two simpler modes remain available
as rollbacks. The first is a weighted average over the present channels,
$\textit{risk}=\sum_{c} w_c s_c / \sum_{c} w_c$ with weights $0.35$, $0.30$, $0.25$, and
$0.10$ for the URL, header, NLP, and attachment channels; its known weakness is that it
\emph{dilutes} a single confident channel, so a URL that the URL engine scored at $100$ with
clean text and headers averages down to roughly $39$ and is undervalued. The second is a
hand-set noisy-OR,
\begin{equation}
\textit{risk} = 100\Big(1 - \prod_{c\in\text{present}}\big(1 - r_c\, s_c/100\big)\Big),
\label{eq:noisyor}
\end{equation}
with per-channel reliabilities $r_c$ defaulting to $1.0$; its single-channel floor preserves a
confirmed signal that the weighted average would dilute. The calibrated probabilistic-OR is
this same OR, except its inputs are per-channel \emph{calibrated} probabilities and not the raw
scores, and
it is the default because calibration makes the channels comparable before they are combined.
The resulting risk maps to four verdict bands, from benign through suspicious and phishing to a
top band reserved for high-confidence phishing or malware.

\subsection{A whole-system benchmark}
Evaluating fusion fairly is harder than evaluating any single engine, because a fusion that
is tuned and tested on the same synthetic generator family will simply memorize that family. We
therefore built a representative whole-system benchmark, a corpus of $10{,}677$ emails in
roughly the $85/5/10$ legitimate/spam/phishing mix of real traffic. Its legitimate, spam, and
real-phishing layers are anchored in public corpora (Enron~\cite{klimt2004enron},
SpamAssassin~\cite{mason2002spamassassin}, Nazario~\cite{nazario2005phishing}, and
Nigerian-419~\cite{radev2008nigerian}, body-only), and a synthetic layer adds full-header,
multi-channel stress cases
that exercise the header and URL channels the body-only real layer cannot. Four hundred rows
are reserved as a leakage-free real-phishing slice. One constraint shapes every number
below: only the NLP channel is the real production model. The URL channel in this benchmark is a
lexical proxy standing in for the fused score of Section~\ref{sec:l2}, and the header channel is a
rule proxy. The benchmark therefore characterizes the \emph{shape} of the fusion decision and
does not amount to a production accuracy claim.

The evaluation protocol is deliberately conservative. Sibling generator families are merged
into mechanism groups, and every email is scored by a calibrated blender that never saw its
group, a pooled out-of-fold scheme that prevents family memorization from masquerading as
generalization. We report the threat view in which phishing is the positive class and both
legitimate and spam are negatives, so that flagging marketing spam counts as a false positive,
since leaving spam in the inbox is intended product behavior. The operating point is a
calibrated phishing probability above $0.255$.

On this corpus the corrected NLP model reaches a three-class accuracy of $0.9518$ on the
natural mix. The fusion comparison in Table~\ref{tab:e2e} is the central result. The shipped
content-bearing blend keeps high recall but at a poor operating point, flagging
\SI{58}{\percent} of spam and reaching an $F_1$ of only $0.774$. Adding the dedicated
phishing-probability head improves precision substantially. Dropping the content-risk channel
altogether, and fusing the URL, header, and phishing-probability channels, gives the best
$F_1$ at $0.914$. The mechanism is straightforward and does not depend on leakage: the content-risk channel equals
one minus the probability of legitimacy, so it fires on real spam, and on held-out real spam
the content-bearing blend mislabels \SI{93.9}{\percent} as phishing while the content-free
blend mislabels only \SI{3.6}{\percent}. Dropping content costs essentially no recall, since
almost no phishing email can be recovered by the content channel alone. With this blender the whole
system realizes nearly all of the model's text-phishing recall: the synthetic adversarial
families are caught in full, and the real Nazario and Nigerian families at $0.976$ and $0.996$.

\begin{table}[t]
\caption{Whole-system fusion on the $10{,}677$-email benchmark (held-out, phishing-vs-rest,
calibrated $P>0.255$). Channels: $u$ URL, $h$ header, $c$ content-risk, $p$
phishing-probability.}
\label{tab:e2e}
\centering
\scriptsize
\begin{tabular}{@{}lrrrr@{}}
\toprule
Fusion & Recall & FPR & $F_1$ & Spam \\
\midrule
Cal.-OR $[u,h,c]$ (shipped)     & 0.898 & 0.044 & 0.774 & 0.58 \\
Cal.-OR $[u,h,c,p]$             & 0.844 & 0.001 & 0.912 & 0.004 \\
Cal.-OR $[u,h,p]$ (recommended) & \textbf{0.881} & 0.005 & \textbf{0.914} & 0.004 \\
\bottomrule
\end{tabular}
\end{table}

A learned fusion was explored and then deliberately rejected, which is a finding in its own right. In a
five-fold comparison at a fixed one-percent false-positive rate, a plain gradient-boosted
stacker~\cite{chen2016xgboost,friedman2001greedy} reached \SI{99}{\percent} recall and a monotonically constrained variant \SI{98.6}{\percent},
against \SI{63}{\percent} for the weighted average. These in-corpus numbers appear strong but are
unreliable: under the mechanism-grouped protocol the learned combiners collapse out of
group, because their apparent advantage derives from memorizing generator families. What generalizes is the
simpler calibrated probabilistic-OR, so that is what we recommend and ship.

Four caveats bound these figures. We give them here, where the results are, and not in a
separate limitations section far from the numbers they qualify. The URL and header channels are proxies. Real phishing families may overlap NLP training data;
degrading their probabilities to the synthetic distribution drops recall from $1.000$ to $0.875$.
The operating threshold was calibrated on synthetic data and shows roughly a twenty-one-percent
false-positive rate on a synthetic-to-real holdout, so it must be recalibrated before deployment.
Live-enrichment timeouts on historical URLs further degrade the proxy URL channel.

\section{Evaluation}
\label{sec:eval}

The detailed result tables sit with their subsystems (Tables~\ref{tab:urlimp},
\ref{tab:nlpres}, \ref{tab:l2dr}, and~\ref{tab:e2e}). This section cross-cuts them against the
service-level objectives; the caveats that bound each claim are consolidated in
Section~\ref{sec:decision} and summarized below.

On detection, the four engines together give a consistent result: strong held-out lexical URL
performance (Section~\ref{sec:urlml}), a fusion sidecar whose detection rate ranges from full
coverage on fresh phishing down to \SI{74.7}{\percent} on subtle generic-domain lures at
\SI{1.3}{\percent} clean benign FPR (Section~\ref{sec:l2}), an NLP classifier that
generalizes to held-out real phishing while passing deployability and metamorphic gates
(Section~\ref{sec:nlp}), and a whole-system fusion reaching $F_1{=}0.914$ on the representative
benchmark (Section~\ref{sec:decision}). The synchronizer sustained $2{,}678$ active indicators
across four feeds at roughly eight seconds per cycle (Section~\ref{sec:ti}).

On latency, every realized engine runs well inside its target. The Domain Guard resolves in tens
of nanoseconds to tens of microseconds, single-URL first-level inference in about $1582$
microseconds, and the NLP model at $9.8$ and $11.3$ milliseconds at the median and 95th
percentile on CPU, an order of magnitude under the per-model service-level targets of five
seconds for URL and ten seconds for NLP at the 99th percentile.

On observability, a single Jaeger trace spanned all services and pipeline-wide Kafka counters
showed 1:1 message conservation (Section~\ref{sec:setup}). This is integration evidence about
message flow, not a detection result.

Several caveats bound these numbers. Per-engine figures pertain to the three realized engines,
not a fully integrated production verdict. Sidecar hard-benign FPR is measured without the
production allowlist; the \SI{1.3}{\percent} clean held-out figure is representative. The
whole-system benchmark uses proxy channels, carries training-contamination risk on its real
phishing slice, and needs threshold recalibration; see Section~\ref{sec:decision} for details.

\section{Discussion}

The results fall into two categories: the three modality engines are benchmarked on real
models and data, while the decision-level fusion is characterized on proxy channels and remains a
preliminary integrated result (Section~\ref{sec:decision}).

The central design principle is modality separation. The NLP model, with URLs and
senders stripped, settles on a text-only Pareto frontier over understated business lures; the Domain
Guard removes easy catastrophic URLs so the sidecar can weight aggressively; and the whole-system
benchmark shows that dropping the content-risk channel (which conflates spam with
phishing) in favor of the dedicated phishing-probability head is what makes the fused verdict
robust to new spam campaigns.

The remaining limitations are tracked, not open: Latin-script URL bias and concept drift;
no synchronizer leader election; fusion thresholds awaiting recalibration; row-level security and
a dead-letter path not yet enforced. The clearest gap is a real, labeled, multi-channel corpus of
full emails with live URLs and headers, the single missing component that would let fusion be retrained on
real production channels in place of today's proxies.

\section{Conclusion}

We have presented a hybrid pipeline for phishing and threat classification whose central claim
is that fusing independent, modality-specific signals beats forcing one model to arbitrate an
entire message. Three engines (URL, NLP, and threat-intelligence synchronization) are
independently benchmarked; a decision-level fusion stage recombines their scores on a
representative whole-system benchmark. The limiting factor is generalization, not
same-distribution accuracy: the NLP redesign lifted held-out real-phishing recall from
\SI{0.8}{\percent} to \SI{87.3}{\percent}, while the URL stack couples deterministic guards with
probabilistic fusion over live enrichment.

The fusion benchmark shows that a calibrated probabilistic-OR over URL, header, and
phishing-probability channels reaches $F_1{=}0.914$ and stays robust to new spam campaigns, but
it uses proxy channels and an operating point awaiting recalibration, a preliminary integrated
result, not a production verdict. Future work acquires the real multi-channel labeled corpus
that would let fusion be validated on production channels, recalibrates thresholds on live
traffic, and replaces hand-tuned cross-modal rules with a learned meta-classifier once that
corpus exists.

\section*{Acknowledgment}
The authors thank Dr.\ Mohamed Samir, the principal academic advisor for this graduation
project, for his guidance and supervision throughout. We are grateful to Zinad, whose
engineers served in an advisory capacity and whose sponsorship supported the work, and to
the members of the thesis defense committee for their evaluation and constructive feedback.

\bibliographystyle{IEEEtran}
\bibliography{references}

@article{chiew2019hefs,
  author  = {Chiew, Kang Leng and Tan, Choon Lin and Wong, KokSheik and Yong, Kelvin S. C. and Tiong, Wei King},
  title   = {A new hybrid ensemble feature selection framework for machine learning-based phishing detection system},
  journal = {Information Sciences},
  volume  = {484},
  pages   = {153--166},
  year    = {2019},
  doi     = {10.1016/j.ins.2019.01.064}
}

@article{tamal2024dataset,
  author  = {Tamal, Maruf Ahmed and Islam, Md. Kabirul and Bhuiyan, Touhid and Sattar, Abdus},
  title   = {Dataset of suspicious phishing {URL} detection},
  journal = {Frontiers in Computer Science},
  volume  = {6},
  pages   = {1308634},
  year    = {2024},
  doi     = {10.3389/fcomp.2024.1308634}
}

@inproceedings{mohammad2012assessment,
  author    = {Mohammad, Rami M. and Thabtah, Fadi and McCluskey, Lee},
  title     = {An Assessment of Features Related to Phishing Websites using an Automated Technique},
  booktitle = {Int. Conf. for Internet Technology and Secured Transactions (ICITST-2012)},
  publisher = {IEEE},
  year      = {2012}
}

@article{potpelwar2025legitphish,
  author  = {Potpelwar, Rachana S. and Kulkarni, Umesh V. and Waghmare, Jaibir M.},
  title   = {{LegitPhish}: A large-scale annotated dataset for {URL}-based phishing detection},
  journal = {Data in Brief},
  volume  = {63},
  pages   = {111972},
  year    = {2025},
  doi     = {10.1016/j.dib.2025.111972}
}

@article{prasad2024phiusiil,
  author  = {Prasad, Arvind and Chandra, Shalini},
  title   = {{PhiUSIIL}: A diverse security profile empowered phishing {URL} detection framework based on similarity index and incremental learning},
  journal = {Computers \& Security},
  volume  = {136},
  pages   = {103545},
  year    = {2024},
  doi     = {10.1016/j.cose.2023.103545}
}

@misc{ciscoumbrella,
  author       = {{Cisco Umbrella}},
  title        = {Cisco Umbrella Popularity List},
  howpublished = {\url{https://s3-us-west-1.amazonaws.com/umbrella-static/index.html}},
  note         = {Daily ranking of most-queried domains on Cisco recursive resolvers}
}

@article{levenshtein1966,
  author  = {Levenshtein, Vladimir I.},
  title   = {Binary codes capable of correcting deletions, insertions, and reversals},
  journal = {Soviet Physics Doklady},
  volume  = {10},
  number  = {8},
  pages   = {707--710},
  year    = {1966}
}

@techreport{edelman2003brand,
  author      = {Edelman, Benjamin},
  title       = {Brand impersonation in domain names},
  institution = {OECD/APWG industry reports on subdomain-based brand abuse},
  year        = {2003}
}

@incollection{platt1999probabilistic,
  author    = {Platt, John C.},
  title     = {Probabilistic outputs for support vector machines and comparisons to regularized likelihood methods},
  booktitle = {Advances in Large Margin Classifiers},
  publisher = {MIT Press},
  year      = {1999}
}

@article{friedman2001greedy,
  author  = {Friedman, Jerome H.},
  title   = {Greedy function approximation: A gradient boosting machine},
  journal = {The Annals of Statistics},
  volume  = {29},
  number  = {5},
  pages   = {1189--1232},
  year    = {2001}
}

@inproceedings{ke2017lightgbm,
  author    = {Ke, Guolin and Meng, Qi and Finley, Thomas and Wang, Taifeng and Chen, Wei and Ma, Weidong and Ye, Qiwei and Liu, Tie-Yan},
  title     = {{LightGBM}: A highly efficient gradient boosting decision tree},
  booktitle = {Advances in Neural Information Processing Systems (NeurIPS)},
  year      = {2017}
}

@article{altan2025dualpath,
  author  = {Altan, Ibrahim and Bachir, Abdulla and Parbhulkar, Yousuf and Rizvi, Abdul Muksith and Farazi, Moshiur},
  title   = {Dual-Path Phishing Detection: Integrating Transformer-Based {NLP} with Structural {URL} Analysis},
  journal = {arXiv preprint arXiv:2509.20972},
  year    = {2025}
}

@article{sajad2026robust,
  author  = {Sajad, U. P.},
  title   = {Explainable Transformer-Based Email Phishing Classification with Adversarial Robustness},
  journal = {arXiv preprint arXiv:2511.12085},
  year    = {2025}
}

@article{tawfik2025xfphishbert,
  author  = {Tawfik, Mohammed and Abu-Ein, Ashraf A. and Abdelhaliem, Ahmed H. and Al-Sharo, Yamin M. and Fathi, Islam S.},
  title   = {Explainable few-shot learning with modern {BERT} for detecting emerging phishing attacks using {XF-PhishBERT}},
  journal = {Scientific Reports},
  volume  = {15},
  year    = {2025},
  doi     = {10.1038/s41598-025-27500-0}
}

@article{safran2025phishinggnn,
  author  = {Safran, Mejdl and Musleh, Abdullah},
  title   = {{PhishingGNN}: Phishing Email Detection Using Graph Attention Networks and Transformer-Based Feature Extraction},
  journal = {IEEE Access},
  volume  = {13},
  pages   = {131390--131399},
  year    = {2025}
}

@article{mendes2025meajor,
  author  = {Mendes, Paulo and Maia, Eva and Pra\c{c}a, Isabel},
  title   = {{MeAJOR} Corpus: A Multi-Source Dataset for Phishing Email Detection},
  journal = {arXiv preprint arXiv:2507.17978},
  year    = {2025}
}

@article{toth2025phish,
  author  = {Toth, Rebeka and Bisztray, Tamas and Gruschka, Nils},
  title   = {The Phish, The Spam, and The Valid: Generating Feature-Rich Emails for Benchmarking {LLM}s},
  journal = {arXiv preprint arXiv:2511.21448},
  year    = {2025}
}

@article{saka2024codebook,
  author  = {Saka, Tarini and Jain, Rachiyta and Vaniea, Kami and K\"{o}kciyan, Nadin},
  title   = {Phishing Codebook: A Structured Framework for the Characterization of Phishing Emails},
  journal = {arXiv preprint arXiv:2408.08967},
  year    = {2024}
}

@article{neshaei2024quantization,
  author  = {Neshaei, Seyed Parsa and Boreshban, Yasaman and Ghassem-Sani, Gholamreza and Mirroshandel, Seyed Abolghasem},
  title   = {The Impact of Quantization on the Robustness of Transformer-based Text Classifiers},
  journal = {arXiv preprint arXiv:2403.05365},
  year    = {2024}
}

@article{ibrahim2024mtl,
  author  = {Ibrahim, Shaymaa and Catal, Cagatay and Kacem, Thabet},
  title   = {The use of multi-task learning in cybersecurity applications: a systematic literature review},
  journal = {Neural Computing and Applications},
  year    = {2024}
}

@article{atanda2026lightweight,
  author  = {Atanda, Oladayo and Aworinde, Halleluyah and van Niekerk, Brett},
  title   = {Lightweight transformer models for scalable phishing email detection: A comparative study of {ALBERT} and {TinyBERT} on a balanced email corpus},
  journal = {Int. J. of Innovative Research and Scientific Studies (IJIRSS)},
  volume  = {9},
  number  = {2},
  pages   = {10--20},
  year    = {2026}
}

@article{sanh2019distilbert,
  author  = {Sanh, Victor and Debut, Lysandre and Chaumond, Julien and Wolf, Thomas},
  title   = {{DistilBERT}, a distilled version of {BERT}: smaller, faster, cheaper and lighter},
  journal = {arXiv preprint arXiv:1910.01108},
  year    = {2019}
}

@inproceedings{kendall2018uncertainty,
  author    = {Kendall, Alex and Gal, Yarin and Cipolla, Roberto},
  title     = {Multi-Task Learning Using Uncertainty to Weigh Losses for Scene Geometry and Semantics},
  booktitle = {Proc. IEEE/CVF Conf. on Computer Vision and Pattern Recognition (CVPR)},
  year      = {2018}
}

@inproceedings{lin2017focal,
  author    = {Lin, Tsung-Yi and Goyal, Priya and Girshick, Ross and He, Kaiming and Doll\'{a}r, Piotr},
  title     = {Focal Loss for Dense Object Detection},
  booktitle = {Proc. IEEE Int. Conf. on Computer Vision (ICCV)},
  year      = {2017}
}

@inproceedings{goodfellow2015adversarial,
  author    = {Goodfellow, Ian J. and Shlens, Jonathon and Szegedy, Christian},
  title     = {Explaining and Harnessing Adversarial Examples},
  booktitle = {Proc. Int. Conf. on Learning Representations (ICLR)},
  year      = {2015}
}

@inproceedings{chen2016xgboost,
  author    = {Chen, Tianqi and Guestrin, Carlos},
  title     = {{XGBoost}: A Scalable Tree Boosting System},
  booktitle = {Proc. ACM SIGKDD Int. Conf. on Knowledge Discovery and Data Mining},
  year      = {2016},
  doi       = {10.1145/2939672.2939785}
}

@article{matthews1975,
  author  = {Matthews, B. W.},
  title   = {Comparison of the predicted and observed secondary structure of {T4} phage lysozyme},
  journal = {Biochimica et Biophysica Acta (BBA) - Protein Structure},
  volume  = {405},
  number  = {2},
  pages   = {442--451},
  year    = {1975},
  doi     = {10.1016/0005-2795(75)90109-9}
}

@article{loshchilov2019adamw,
  author  = {Loshchilov, Ilya and Hutter, Frank},
  title   = {Decoupled Weight Decay Regularization},
  journal = {arXiv preprint arXiv:1711.05101},
  year    = {2019},
  note    = {Published at ICLR 2019}
}

@inproceedings{broder1997minhash,
  author    = {Broder, Andrei Z.},
  title     = {On the resemblance and containment of documents},
  booktitle = {Proc. Compression and Complexity of Sequences (SEQUENCES'97)},
  pages     = {21--29},
  year      = {1997},
  doi       = {10.1109/SEQUEN.1997.666900}
}

@inproceedings{manku2007simhash,
  author    = {Manku, Gurmeet Singh and Jain, Arvind and Sarma, Anish Das},
  title     = {Detecting near-duplicates for web crawling},
  booktitle = {Proc. Int. Conf. on World Wide Web (WWW)},
  year      = {2007},
  doi       = {10.1145/1242572.1242592}
}

@book{pearl1988,
  author    = {Pearl, Judea},
  title     = {Probabilistic Reasoning in Intelligent Systems: Networks of Plausible Inference},
  publisher = {Morgan Kaufmann},
  address   = {San Mateo, CA},
  year      = {1988}
}

@misc{publicsuffix,
  author       = {{Mozilla Foundation}},
  title        = {Public Suffix List},
  howpublished = {\url{https://publicsuffix.org/}},
  note         = {Community-maintained list of public suffixes for registrable-domain extraction}
}

@misc{phishtank,
  author       = {{Cisco Talos Intelligence Group}},
  title        = {PhishTank},
  howpublished = {\url{https://data.phishtank.com/}},
  note         = {Community-verified phishing URL feed}
}

@misc{openphish,
  author       = {{OpenPhish}},
  title        = {OpenPhish Phishing Intelligence},
  howpublished = {\url{https://openphish.com/}},
  note         = {Commercial phishing URL feed}
}

@misc{urlhaus,
  author       = {{abuse.ch}},
  title        = {URLhaus Malware URL Exchange},
  howpublished = {\url{https://urlhaus.abuse.ch/}},
  note         = {Community malware URL feed}
}

@misc{threatfox,
  author       = {{abuse.ch}},
  title        = {ThreatFox {IOC} Sharing Platform},
  howpublished = {\url{https://threatfox.abuse.ch/}},
  note         = {Community indicator-of-compromise sharing}
}

@misc{malwarebazaar,
  author       = {{abuse.ch}},
  title        = {MalwareBazaar},
  howpublished = {\url{https://bazaar.abuse.ch/}},
  note         = {Malware sample and hash exchange}
}

@misc{akshatsharma2023spamhamphish,
  author       = {Sharma, Akshat},
  title        = {The Biggest Spam Ham Phish Email Dataset ({300000+})},
  howpublished = {\url{https://www.kaggle.com/datasets/akshatsharma2/the-biggest-spam-ham-phish-email-dataset-300000}}
}

@misc{tanvirahmed2024eduphish,
  author       = {Ahmed, Tanvir},
  title        = {Education-Targeted Phishing Email Dataset},
  howpublished = {\url{https://www.kaggle.com/datasets/tanvirahmed0981/education-targeted-phishing-email-dataset}}
}

@inproceedings{greco2024llmphish,
  author    = {Greco, Francesco and Desolda, Giuseppe and Esposito, Andrea and Carelli, Alessandro},
  title     = {David versus {Goliath}: Can Machine Learning Detect {LLM}-Generated Text? A Case Study in the Detection of Phishing Emails},
  booktitle = {Proc. Italian Conf. on CyberSecurity (ITASEC 2024)},
  series    = {CEUR Workshop Proceedings},
  volume    = {3731},
  year      = {2024},
  url       = {https://ceur-ws.org/Vol-3731/paper41.pdf}
}

@misc{nazario2005phishing,
  author       = {Nazario, Jose},
  title        = {Nazario Phishing Corpus},
  howpublished = {\url{https://monkey.org/~jose/phishing/}},
  year         = {2005},
  note         = {CC-BY-4.0}
}

@misc{radev2008nigerian,
  author       = {Radev, Dragomir},
  title        = {{CLAIR} Collection of Fraud Email},
  howpublished = {ACL Data and Code Repository, ADCR2008T001},
  year         = {2008},
  url          = {https://www.aclweb.org/aclwiki/CLAIR_collection_of_fraud_email_(Repository)}
}

@misc{puyang2025seven,
  author       = {Puyang},
  title        = {Seven Phishing Email Datasets},
  howpublished = {\url{https://huggingface.co/datasets/puyang2025/seven-phishing-email-datasets}},
  year         = {2025}
}

@inproceedings{klimt2004enron,
  author    = {Klimt, Bryan and Yang, Yiming},
  title     = {The {Enron} Corpus: A New Dataset for Email Classification Research},
  booktitle = {European Conference on Machine Learning},
  pages     = {217--226},
  year      = {2004},
  doi       = {10.1007/978-3-540-30115-8_22}
}

@misc{mason2002spamassassin,
  author       = {Mason, Justin},
  title        = {SpamAssassin Public Mail Corpus},
  howpublished = {\url{https://spamassassin.apache.org/old/publiccorpus/}},
  year         = {2002}
}

\end{document}